\documentclass{Interspeech}

\interspeechcameraready

\title{Exploiting Context-dependent Duration Features for  Voice Anonymization Attack Systems}

\author[affiliation={1}]{Natalia}{Tomashenko}
\author[affiliation={1}]{Emmanuel}{Vincent}
\author[affiliation={2}]{Marc}{Tommasi}

\affiliation{Université de Lorraine, CNRS, Inria, LORIA}{F-54000, Nancy}{France}
\affiliation{Université de Lille, CNRS, Inria, Centrale Lille}{UMR 9189 - CRIStAL, Lille}{France}
\email{natalia.tomashenko@inria.fr, emmanuel.vincent@inria.fr, marc@tommasi@inria.fr}

\usepackage{comment}
\usepackage{numprint}
\usepackage{makecell}
\usepackage{multirow}
\usepackage{rotating}
\usepackage{cite}
\usepackage{colortbl,dcolumn}
\usepackage[dvipsnames]{xcolor}

\RequirePackage{etoolbox}
\RequirePackage{pgf} 
\RequirePackage{xcolor}

\newcommand{\gradientcelld}[8]{
\xdef\lowvalx{#2}%
\xdef\midvalx{#3}%
\xdef\maxvalx{#4}%
\xdef\lowcolx{#5}%
\xdef\midcolx{#6}%
\xdef\highcolx{#7}%
\xdef\opacityx{#8}%

\ifdimcomp{#1pt}{>}{\maxvalx pt}{\cellcolor{\highcolx!100.0!\midcolx!\opacityx}#1}{
\ifdimcomp{#1pt}{<}{\midvalx pt}{%
\ifdimcomp{#1pt}{<}{\lowvalx pt}{\cellcolor{\midcolx!0.0!\lowcolx!\opacityx}#1}{
     \pgfmathparse{int(round(100*(#1/(\midvalx-\lowvalx))-(\lowvalx*(100/(\midvalx-\lowvalx)))))}%
    \xdef\tempa{\pgfmathresult}%
    \cellcolor{\midcolx!\tempa!\lowcolx!\opacityx}#1%
}}{
     \pgfmathparse{int(round(100*(#1/(\maxvalx-\midvalx))-(\midvalx*(100/(\maxvalx-\midvalx)))))}
    \xdef\tempb{\pgfmathresult}%
    \cellcolor{\highcolx!\tempb!\midcolx!\opacityx}#1%
}}
}

\definecolor{darkcyan}{rgb}{0.0, 0.55, 0.55}

\newcommand{\g}[1]{\gradientcelld{#1}{1.8}{22}{42}{YellowOrange}{white}{darkcyan}{70}}

\begin{document}

\maketitle

\begin{abstract}  
The temporal dynamics of speech, encompassing variations in rhythm, intonation, and speaking rate, contain important and  unique information about speaker identity. 
This paper proposes a new method for representing speaker characteristics by extracting context-dependent duration embeddings from speech temporal dynamics.  We develop novel attack models using these representations and  analyze the potential vulnerabilities in speaker verification and voice anonymization systems.The experimental results show that the developed attack models provide a significant improvement in speaker verification performance for both  original and anonymized data in comparison with simpler representations of speech temporal dynamics reported in the literature. 
\end{abstract}

\section{Introduction}

The widespread use of speech data today raises major privacy concerns, leading to its protection under the scope of privacy regulations like the European General Data Protection Regulation (GDPR)~ \cite{nautsch2019gdpr}.
Voice recordings contain a large amount of personal information, and beyond  revealing a speaker's identity, speech data can disclose other sensitive characteristics including age, gender, health conditions,  emotional state, personality traits, ethnic background, and socioeconomic status. 
Understanding and protecting this vast amount of personal information is becoming increasingly important as voice technology becomes more prevalent in our daily lives.

A common method for preserving the privacy of speech data is voice anonymization which aims to suppress personally identifiable characteristics of the speaker, while preserving the linguistic and paralinguistic content \cite{tomashenko2020introducing}.
Voice anonymization techniques have evolved significantly. They include two broad categories of methods.

The first category, signal processing-based methods, employs simple signal transformations to alter voice characteristics. These include spectral warping utilizing the McAdams coefficient \cite{patino2020speaker}, pitch shifting through time-scale modification \cite{mawalim22_spsc}, and others \cite{gupta2020designn,tavi2022improving}, offering straightforward but limited anonymization capabilities.

The second and more complex category comprises neural voice conversion based methods \cite{fang2019speaker,miao2023speaker,srivastava2021,champion2023,yao2024musa,miao2022language,yao2024npu,saini2023speaker,webber2024voice}, 
that operate by  disentangling various speech attributes --- including content, speaker characteristics, pitch, and emotion --- before selectively anonymizing specific attributes and reconstructing the speech signal using  speech synthesis models.
Most state-of-the-art voice conversion based anonymization systems typically leverage large-scale pre-trained models for  attribute extraction, demonstrating superior performance in both content preservation and privacy protection compared to signal processing-based methods.

While researchers have made significant advances in voice anonymization techniques and conducted several studies on the speaker information carried by pitch \cite{tavi2022improving,srivastava2021,champion2023}, little attention has been paid to the role of speech temporal dynamics in voice anonymization.
Many state-of-the-art anonymization systems keep speech rate and phoneme durations unchanged~\cite{
fang2019speaker,srivastava2021,miao2022language,panariello_speaker_2023}.
Cascaded automatic speech recognition (ASR) and text-to-speech (TTS) systems \cite{sinha2022eli,xinyuan2024hltcoe} present rare exceptions, where word-level or phoneme-level transcripts from ASR are used by TTS to synthesize linguistic content with a new target voice. Although these systems likely avoid retaining speaker identity, they typically fail to preserve the paralinguistic attributes crucial for real-life voice anonymization applications.

Prior research relevant to our study~\cite{prajapati2022voice} involves using speed perturbation with a constant factor as an anonymization technique, either independently or combined with anonymization based on a cycle consistent generative adversarial network (CycleGAN). These studies demonstrated that speech rate perturbation with a constant factor reduces the effectiveness of automatic speaker verification systems against both \textit{ignorant} and \textit{lazy-informed} attackers. However, they did not account for the more robust \textit{semi-informed} attack model that is now the standard~\cite{tomashenko2024first}.

The studies in \cite{fujita21_interspeech,fujita2024speech} proposed speech rhythm-based speaker embeddings for duration modeling in multi-speaker speech synthesis.
Another recent study \cite{tomashenko2025analysis}, using a simple approach based on comparison of average phoneme duration vectors, has revealed the importance of speech temporal dynamics analysis in voice anonymization research.

This paper builds upon  \cite{tomashenko2025analysis,fujita2024speech} and makes a further step in this direction, performing a more systematic and advanced analysis of the speaker information  conveyed in speech temporal dynamics. 
We consider 
two
levels of information regarding  speech temporal dynamics:  
statistics of average phoneme durations in utterances and complete set of phoneme durations in the given utterances. Based on the corresponding representations, we propose speaker verification models that can be used to attack voice anonymization systems.
A key contribution is to propose a new method to encode the complete set of phoneme durations in the form of learned context-dependent duration embeddings and to show the strong improvement in attack performance achieved with this representation.

Section~\ref{sec:represnt-of-speech} describes the raw duration feature sequences and Section~\ref{sec:attacks} explains how context-dependent duration embeddings can be extracted and used to construct attacks. Sections~\ref{sec:setup} and \ref{sec:results} describe the experimental setup and the results. Section~\ref{sec:concl} concludes the paper.

\begin{figure*}[t]
  \centering
  \includegraphics[width=\linewidth]{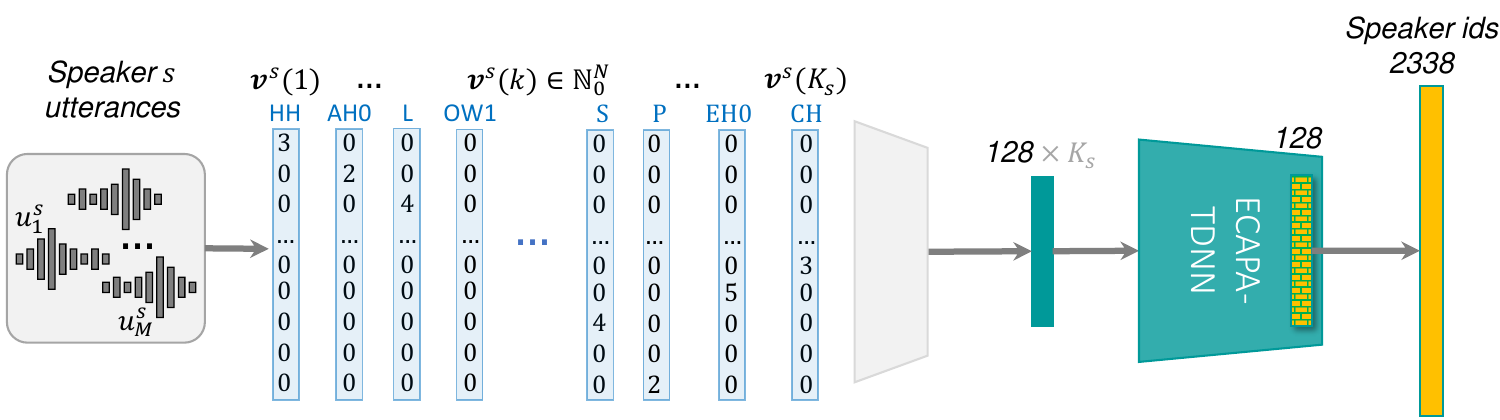}
  \caption{Attack model training on raw duration feature sequences, yielding a 128-dimensional context-dependent duration embedding (yellow).}
  \label{fig:cdd}
\end{figure*}

\section{Raw duration feature sequences}
\label{sec:represnt-of-speech}

Previous works \cite{bulgakova2015speaker,tomashenko2025analysis} show that speaking rate and average durations of phonemes contain a significant amount of speaker information.
In this study, our objective is to discover representations that more efficiently capture speaker information conveyed by speech temporal dynamics. We also aim to better understand how much speaker information can be retrieved from these different representations.

To do so, we consider raw duration feature sequences that are more informative than average phoneme durations
and include all information about individual phoneme occurrences
in the given utterance(s), including their context. 
Let us denote by $N$ the number of phoneme classes 
$ph_1,\ldots,ph_N$, 
and by $K_s$ the total number of phones
$phone(1),\ldots,phone(K_s)$
in the considered utterance(s) of speaker $s$.
For $k\in[1,K_s]$, each phone $phone(k)$ in the utterance(s) of speaker $s$ is represented by a corresponding $N$-dimensional vector 
$\textbf{\textit{v}}^s(k)\in\mathbb{N}_0^N$: $\textbf{\textit{v}}^s(k)=[v^s_1(k),\ldots,v^s_N(k)]$ --
a one-hot vector multiplied by the phone length (in acoustic frames) $len(phone(k))$, where the index $i$ of a non-zero component
$v_i^s(k)$ 
corresponds to the actual phoneme class $ph_i$: $phone(k)\in ph_i$. In other words,  
\begin{equation}
    v^s_i(k)=
    \begin{cases}
    len(phone(k)), phone(k) \in ph_i \\
    0, phone(k) \notin ph_i
    \end{cases}.
\end{equation}
The phone lengths are extracted from the speech utterances using phonetic alignment.

\section{Context-dependent duration embeddings and attack model}
\label{sec:attacks}
 
The raw duration feature sequences proposed in Section~\ref{sec:represnt-of-speech}
allow us to build an efficient speaker verification model that learns a context-dependent duration embedding for the given utterance(s) of each speaker. This model is expected to be robust to modifications of the speech signal during anonymization other than changes in the temporal dynamics. Therefore, it can be used by attackers against voice anonymization.

The training process is illustrated in Figure~\ref{fig:cdd}.
To reduce overfitting, instead of training the model on utterances, we train it on chunks of variable length ($32-256$ phones) with a random shift in the first utterance $u_1$ in every chunk, i.e. 
we remove a random number of phones $r$ from the first utterance of each chunk. This random variable follows a uniform distribution $r\sim U(0, \min\{len(u_1), len(chunk)\})$, where $len(u_1)$ represents the  number of phones in the first utterance $u_1$, and $len(chunk)$ is  number of phones in the entire chunk.
We adopt the standard ECAPA-TDNN (emphasized channel attention, propagation and aggregation in the time delay neural network) architecture
\cite{desplanques2020ecapa} and use 128-dimensional projected versions of the raw duration feature vectors as inputs.
A speaker embedding which encodes the context-dependent duration information is extracted from the last fully connected layer of the ECAPA-TDNN shown in yellow. The output layer performs speaker classification. The projection layer and the ECAPA-TDNN are jointly trained with a cross-entropy loss.

At test time, speaker embeddings are computed for the trial utterance(s) and the enrollment speaker using the trained network, and automatic speaker verification (ASV) scores are obtained using the cosine distance between embeddings.

\section{Experimental setup}
\label{sec:setup}

\subsection{Data}

Experiments were conducted on the \textit{LibriSpeech}\footnote{\label{fn:url1}\textit{LibriSpeech}: \url{http://www.openslr.org/12}}~\cite{panayotov2015librispeech} corpus of read English audiobooks.
It contains approximately 1,000~hours of speech from 2,484 speakers sampled at 16~kHz.
The training data is the \textit{LibriSpeech-train-960} subset with 2,338 speakers.

For development and evaluation we used the , 
\textit{dev-clean} 
and \textit{test-clean} subsets of \textit{LibriSpeech}.
For experiments on anonymized data, we used 
the \textit{dev-clean} and \textit{test-clean} subsets anonymized by a
voice anonymization system
as described in  Section~\ref{sec:anonym_systems}.

To perform phonetic alignment, two triphone Gaussian mixture model - hidden Markov model (GMM-HMM) acoustic models were trained using the Kaldi speech recognition toolkit \cite{povey2011kaldi} on the following training data: (1)~original   \textit{LibriSpeech-train-960}; and
(2)~\textit{LibriSpeech-train-clean-360} anonymized by the voice anonymization system. 
The second model was used to perform segmentation of the anonymized development and evaluation datasets. 

 In our experiments, we utilize exact text transcripts to obtain phonetic alignment. However, in real-world application scenarios, these alignments can be automatically obtained by ASR. The aim of this paper is to prove the importance of speech temporal
 dynamics for voice anonymization against a strong attacker, which is a necessary step before considering real-world application scenarios possibly involving weaker attacks. The motivation for using exact transcripts is therefore to eliminate, as much
as possible, all factors that may negatively impact the attacker’s performance. These include transcription errors caused by ASR.

\begin{table*}[h]
    \centering
    \caption{Automatic speaker verification results (EER,$\%$) on original and anonymized data depending on the number of utterances ($\#$utter) used to compute speaker embeddings, for three different attack models.}
    \begin{tabular}{cccccccc}
        \toprule
        \multicolumn{2}{c}{}  &  \multicolumn{2}{c}{\textbf{original data}} & \multicolumn{3}{c}{\textbf{anonymized data}} \\
        \cmidrule(lr){3-4} \cmidrule(lr){5-7} 
        & & \textbf{A1} & \textbf{A2} & \textbf{A1} & \textbf{A2} & \textbf{A3} \\
      \textbf{data} & $\#$ \textbf{utter}  & \textbf{metric} & \textbf{\makecell{duration \\ features}} & \textbf{metric} & \makecell{\textbf{\makecell{duration \\ features}}} & \textbf{fbanks} \\
        \midrule
      &  1 & \g{38.2} {\textcolor{gray}{\footnotesize{$\pm0.7$}}} &  \g{24.8} {\textcolor{gray}{\footnotesize{$\pm0.6$}}} & \g{41.7} {\textcolor{gray}{\footnotesize{$\pm0.7$}}}& \g{32.4} {\textcolor{gray}{\footnotesize{$\pm0.6$}}} & \g{25.9} {\textcolor{gray}{\footnotesize{$\pm0.6$}}} \\

     dev-clean &  4 & \g{32.2} {\textcolor{gray}{\footnotesize{$\pm0.6$}}} & \g{ 7.6} {\textcolor{gray}{\footnotesize{$\pm0.4$}}} & \g{32.1} {\textcolor{gray}{\footnotesize{$\pm0.6$}}} & \g{12.5} {\textcolor{gray}{\footnotesize{$\pm0.5$}}}& \g{ 7.8} {\textcolor{gray}{\footnotesize{$\pm0.4$}}} \\
  
      &  8 & \g{25.9} {\textcolor{gray}{\footnotesize{$\pm0.6$}}} & \g{ 2.8} {\textcolor{gray}{\footnotesize{$\pm0.2$}}} & \g{26.1} {\textcolor{gray}{\footnotesize{$\pm0.6$}}}& \g{ 6.0} {\textcolor{gray}{\footnotesize{$\pm0.3$}}} & \g{ 3.2} {\textcolor{gray}{\footnotesize{$\pm0.2$}}} \\
    
        \midrule
      &  1 & \g{41.9} {\textcolor{gray}{\footnotesize{$\pm0.7$}}} & \g{24.2} {\textcolor{gray}{\footnotesize{$\pm0.6$}}}& \g{41.8} {\textcolor{gray}{\footnotesize{$\pm0.7$}}}& \g{30.0} {\textcolor{gray}{\footnotesize{$\pm0.6$}}}& \g{22.9} {\textcolor{gray}{\footnotesize{$\pm0.6$}}}\\
   
    test-clean  &  4 & \g{32.2} {\textcolor{gray}{\footnotesize{$\pm0.6$}}}& \g{ 5.9} {\textcolor{gray}{\footnotesize{$\pm0.3$}}} & \g{31.8} {\textcolor{gray}{\footnotesize{$\pm0.6$}}}& \g{10.4} {\textcolor{gray}{\footnotesize{$\pm0.4$}}}& \g{ 5.4} {\textcolor{gray}{\footnotesize{$\pm0.3$}}} \\
   
      &  8 & \g{26.0} {\textcolor{gray}{\footnotesize{$\pm0.6$}}} & \g{ 1.8} {\textcolor{gray}{\footnotesize{$\pm0.2$}}} & \g{25.9} {\textcolor{gray}{\footnotesize{$\pm0.6$}}} & \g{ 3.9} {\textcolor{gray}{\footnotesize{$\pm0.3$}}} & \g{ 2.0} {\textcolor{gray}{\footnotesize{$\pm0.2$}}} \\
      
        \bottomrule
    \end{tabular}
    \label{tab:res}
\end{table*}

\subsection{Anonymization system}
\label{sec:anonym_systems}

 To investigate the impact of voice anonymization, as an example, we consider a state-of-the-art speaker voice anonymization system that keeps the original temporal phoneme dynamics unchanged, but modifies speaker identity and some prosodical characteristics such as pitch and energy.
This voice anonymization system, proposed in \cite{meyer2023prosody}
and used as baseline \textbf{B3} in the VoicePrivacy 2024 Challenge \cite{tomashenko2024voiceprivacy}, is uses phonetic transcription and a generative adversarial network (GAN) that generates artificial pseudo-speaker embeddings. Anonymization is performed in three steps: 

\begin{enumerate}
    \item 

extraction of the speaker embedding, phonetic transcription, pitch,
energy, and phone duration from the original audio waveform; 

    \item 

speaker embedding anonymization, pitch and energy modification; 

    \item 

synthesis of an anonymized speech waveform from the anonymized speaker embedding, modified pitch and energy features, original phonetic transcripts and original phone durations. 

\end{enumerate}

The automatic speaker verification results in terms of equal error rate (EER)  on the \textit{LibriSpeech} test set for anonymized data, according to \cite{tomashenko2024voiceprivacy}, are around $27-28\%$ for the strongest \textit{semi-informed} attacker trained on \textit{utterance-level} anonymized data.

\subsection{Attack model configuration}
\label{sec:attack-experiments}

We perform experiments with the attack models described in Section~\ref{sec:attacks}.
In all the reported experiments, we use the ARPAbet symbol set corresponding to the Carnegie Mellon University pronunciation dictionary\footnote{\url{http://www.speech.cs.cmu.edu/cgi-bin/cmudict}} with 
$N=336$ 
phoneme classes that take into account position in the word and stress. 

The configuration and parameters of the ECAPA-TDNN are the same as those used in the \textit{SpeechBrain} recipe\footnote{\url{https://github.com/speechbrain/speechbrain/tree/develop/recipes/VoxCeleb/SpeakerRec}}.
The size of the speaker embedding extracted from the ECAPA-TDNN models and used for ASV score computation is $128$. 
The model is trained using the cross-entropy criterion.

\subsection{Baseline models}
\label{sec:baselines}
In the following, we use two models for comparison.

As the first baseline model,  we consider 
the simple metric-based approach in \cite{tomashenko2025analysis}. We  shortly review it in this section.
 For a speaker $s$ with utterances $u^s_{1},\ldots, u^s_{M}$, we define an $N$-dimensional vector ${\boldsymbol{\mu}}^{s}=[\mu^{s}_{1},\ldots,\mu^{s}_{N}]\in\mathbb{R}_{\geq0}^N$. This vector represents the average durations of phonemes $ph_1,\ldots,ph_N$ calculated across all utterances from speaker $s$. In cases where a particular phoneme is absent from the speaker's utterances, we fill in the vector with approximated values to ensure completeness. These approximations are derived from the mean duration of all phonemes present in the considered utterances for that speaker. Through this approach, any utterance or set of utterances from speaker $s$ can be  represented by the vector ${\boldsymbol{\mu}}^{s}$. 
The ASV score is computed using the metric proposed in \cite{tomashenko2025analysis}: 
\begin{equation}
\rho(s_i,s_j)=1-\frac{1}
{N}\sum_{n=1}^{N}\min\left\{\frac{\mu^{s_i}_n}{\mu^{s_j}_n}, \frac{\mu^{s_j}_n}{\mu^{s_i}_n}\right\},
\end{equation}
where $\mu^{s_i}_n$, $\mu^{s_j}_n$ are the $n$-th coordinates of mean duration vectors  $\boldsymbol{\mu}^{s_i}$, $\boldsymbol{\mu}^{s_j}$  for speakers $s_i$, $s_j$.

As a second baseline model, we use an ASV model trained on anonymized data on conventional 80-dimensional filter bank features. This model is similar to the \textit{semi-informed} attack model used in the VoicePrivacy 2024 Challenge evaluation setup \cite{tomashenko2024voiceprivacy}. The configuration and parameters of the ECAPA-TDNN are the same as those used in the \textit{SpeechBrain} \textit{VoxCeleb} recipe.

\section{Results}
\label{sec:results}

The results on the original evaluation data for different attack models are summarized  in Table~\ref{tab:res} in terms of equal error rate (EER, $\%$). The  EER values are reported with   $95\%$ confidence intervals, calculated as suggested in \cite{bengio2004statistical}.
In this table, the results are given for development and evaluation data, on the original unprocessed (the third and fourth columns) and anonymized (the last three columns) data for three different attack models:

\begin{enumerate}

    \item \textbf{A1. metric} -- the baseline metric-based approach \cite{tomashenko2025analysis} described in Section~\ref{sec:baselines}

    \item \textbf{A2. duration features} -- the proposed attack model relying on learned context-dependent duration embeddings

    \item  \textbf{A3. fbanks} (only for anonymized data) -- a \textit{semi-informed} attack model trained on anonymized data using conventional filter bank features.

\end{enumerate}

It's important to note that models trained using only temporal dynamics information do not require retraining when used with anonymized data.
This is one advantage of the proposed model with context-dependent duration embeddings which, unlike the conventional \textit{semi-informed} attack model, is trained on original data and does not require retraining on anonymized data. Thus, a single universal attack model can be applied against different anonymization algorithms.

Table~\ref{tab:res} shows the results for different numbers of utterances used to compute the speaker embeddings. 
In these experiments we use the same number of utterances for enrollment and trials and consider three different setups: $1$, $4$, and $8$ utterances. 
The ASV performance improves significantly with increasing number of utterances, especially for the model using on context-dependent duration embeddings and for the \textit{semi-informed} attack model trained on anonymized data on filter bank features. On original test data, for \textbf{A2}, the EER reduces from $24.8\%$ to $2.8\%$ when the number of utterances increases from $1$ to $8$.

The new  attack model \textbf{A2} significantly outperforms
the  baseline metric-based attack system \textbf{A1}. On the original test data, the  EER reduction for \textbf{A1} w.r.t. \textbf{A2} is $8-26\%$ 	absolute ($42-93\%$ relative), while on the anonymized test data	the  EER reduction is	$12-22\%$ absolute and 
$28-85\%$  relative.

We can observe some degradation of the results for \textbf{A2} on anonymized data w.r.t. the results on original data in all cases, unlike for the  baseline model \textbf{A1}, for which the results on the original and anonymized data are similar. However, \textbf{A2} substantially outperforms the baseline \textbf{A1} on the anonymized data as well.

Many automatic speaker verification systems typically utilize multiple utterances per speaker for enrollment and a single trial utterance for each test comparison. Table~\ref{tab:2} demonstrates the performance of the models close to this scenario: for the \textbf{A2} model on original and anonymized data, and for \textbf{A3} on anonymized data. In these experiments, the number of enrollment utterances is either 8 or 16, and a single test trial utterance is used.

 The results for both the \textbf{A2} and \textbf{A3} models demonstrate that we can achieve a relatively low EER with a small number of utterances for enrollment and trials.

\begin{table}[h]
    \centering
    \caption{Automatic speaker verification results (EER,$\%$)  on original and anonymized data for different attack models and varying number of utterances for enrollment (enr), with one utterance for trial (trl).}
    \label{tab:2}
    \scalebox{0.95}{
    \begin{tabular}{ccccc}
        \toprule
       \multicolumn{2}{c}{}    &   \textbf{original data} & \multicolumn{2}{c}{\textbf{anonymized data}} \\ \cmidrule(lr){3-3} \cmidrule(lr){4-5}
         & & \textbf{A2} & \textbf{A2} & \textbf{A3} \\
         & \textbf{ $\#$ utter } & \textbf{duration} & \textbf{duration} &  \\
      \textbf{data}   & \textbf{(enr,trl)} & \textbf{features} & \textbf{features} & \textbf{fbanks}\\
        \midrule
        \multirow{2}{*}{dev-clean} & (8, 1) & \g{15.3} & \g{21.5} & \g{17.1} \\
                  & (16,1) & \g{14.5} & \g{20.7} & \g{16.5} \\  \midrule
        \multirow{2}{*}{test-clean} & (8, 1) & \g{14.5} & \g{18.2} & \g{14.1} \\
                   & (16,1) & \g{14.0} & \g{16.6} & \g{13.1} \\
        \bottomrule
    \end{tabular}
    }
\end{table}

\section{Conclusions}
\label{sec:concl}

Our introduction of context-dependent duration embeddings provides new possibilities for robust speaker verification and privacy-preserving voice technologies.
The proposed context-dependent duration embeddings and attack model allowed us to significantly outperform 
the results reported in the literature for simpler representations of speech temporal dynamics  \cite{tomashenko2025analysis}, providing  $8-26\%$ absolute ($42-93\%$ relative) EER reduction on original unprocessed data, and  $12-22\%$ absolute  
($28-85\%$  relative) EER reduction on anonymized data.
This model offers a key advantage: unlike traditional \textit{semi-informed} attack models \cite{tomashenko2025first}, it can be trained directly on original data, eliminating the need for retraining on anonymized data.
The results of this study can provide guidance for improving current voice anonymization algorithms.

Future research directions include extension of the study to other speech corpora, 
investigation of the complementarity of the proposed context-dependent duration embeddings and conventional speaker embeddings,
and development of more advanced attack models
and countermeasures that
can conceal speaker temporal patterns effectively and withstand new attack scenarios.

\section{Acknowledgements}
This work was supported by the French National Research Agency under project Speech Privacy and project IPoP of the Cybersecurity PEPR.  Experiments were carried out using the Grid’5000 testbed.

\bibliographystyle{IEEEtran}
\bibliography{mybib}

\end{document}